\documentclass[galaxies,article,accept,pdflatex,oneauthor]{Definitions/mdpi} 

\firstpage{1} 
\pubvolume{1}
\issuenum{1}
\articlenumber{0}
\pubyear{2025}
\copyrightyear{2025}
\datereceived{ } 
\daterevised{ } 
\dateaccepted{ } 
\datepublished{ } 
\hreflink{https://doi.org/} 

\makeatletter
\@ifundefined{endpage}{\def\endpage#1{} }{}
\makeatother

\makeatletter
\let\@\,  
\makeatother

\usepackage{color}
\usepackage{soul}

\providecommand{\tcdegree}{\ensuremath{^\circ}}

\Title{Using Light Curve Derivatives to Estimate the Fill-out Factor of Overcontact Binaries}
\TitleCitation{Fill-out factor estimates for overcontact binaries}

\Author{Shinjirou Kouzuma $^{1}$}
\AuthorNames{Shinjirou Kouzuma}
\AuthorCitation{Kouzuma, S.}

\address{%
$^{1}$ \quad Faculty of Liberal Arts and Sciences, Chukyo University, 101-2 Yagoto-honmachi, Showa-ku, Nagoya, Aichi 466-8666, Japan; skouzuma@lets.chukyo-u.ac.jp\\
}

\corres{Correspondence: skouzuma@lets.chukyo-u.ac.jp}

\abstract{
We propose a simple method for estimating the fill-out factor of overcontact binary systems using the derivatives of light curves. 
We synthesized 74,431 sample light curves, covering the typical parameter space of overcontact binaries. 
On the basis of a recent study that proposed a new classification scheme using light curve derivatives up to the fourth order, the sample light curves were classified. 
Among the classified types, for systems exhibiting high mass ratios and high inclinations (i.e., SPf type), we found that the fill-out factor has a strong correlation with the time interval between two local extrema in the third derivatives of their light curves. 
An empirical formula to estimate the fill-out factor was derived using regression analysis for the identified correlation. 
Application to real overcontact binary data demonstrated that the proposed method is practical for obtaining reliable estimates of the fill-out factor and its associated uncertainties. 
}
\keyword{eclipsing binaries; overcontact binaries; fill-out factor: estimation method
} 

\begin{document}

\section{Introduction}
Overcontact binaries are close binary systems in which both stars overfill their Roche lobes. 
It is widely accepted that overcontact binaries evolve from detached systems with short orbital periods (e.g., \citet{Demircan2006-MNRAS} and references therein). 
Previous studies have suggested that the merging of overcontact binaries can result in the formation of blue stragglers \citep{Rasio1995-ApJ,Yildiz2014-MNRAS}. 
Therefore, understanding the properties and evolution of overcontact binaries provides valuable insights into the evolutionary pathways of close binary systems and helps in investigating their ultimate fates. 

The fill-out factor is a binary parameter that quantifies the degree of overfill (overcontact). 
Although several researchers have proposed different definitions of the fill-out factor \citep{Mochnacki1972-MNRAS,Rucinski1973-AcA,Yamasaki1975-PASJ,Lucy1979-ApJ}, recent studies, including synthetic codes such as Binary Maker \citep{Bradstreet2005-SASS}, Djura{\v s}evi{\'c}'s code \citep{Djurasevic1998-AAS}, and PHOEBE \citep{Prsa2016-ApJS}, generally adopt the following form: 
\begin{equation}
	f = \frac{\Omega - \Omega_\text{in}}{\Omega_\text{out} - \Omega_\text{in}}, 
\end{equation}
where $\Omega$, $\Omega_\text{in}$, and $\Omega_\text{out}$ are the potentials of the stellar surface, inner and outer Roche lobes, respectively. 
The inner and outer Roche lobes are the inner and outer Lagrangian zero-velocity equipotential surfaces crossing the Lagrange points L1 and L2 , respectively \citep{Mochnacki1981-ApJ,Stepien2015-AA}. 
This factor expresses the stellar surface potential as a fraction of the total potential difference between the inner and outer critical surfaces. 

The fill-out factor is expected to be closely related to the evolution of overcontact binaries. 
In the evolution of close binary systems, including overcontact binaries, mass exchange between the component stars and mass loss from the system through the L2 point are crucial (e.g., \citet{Paczynski1971-ARAA, Kuiper1941-ApJ}). 
Such mass transfer processes can occur when a component star fills or overfills its Roche lobe. 
Accordingly, the fill-out factor should be closely associated with the properties of mass transfer. 
Many studies have discussed the impact of the fill-out factor on the evolution of overcontact binaries \citep{Rasio1995-ApJ887,Zhou2015-AJ, Wadhwa2021-MNRAS,Pesta2023-AA}. 
Therefore, understanding and accurately estimating the fill-out factor is essential for interpreting the evolution of overcontact binaries and their observational signatures. 

Light curve (LC) analysis is generally required for determining the fill-out factor. 
In this analysis, appropriate initial parameter settings and iterative methods are required to obtain convergent and reliable solutions. 
Binary parameters are often determined using optimization techniques. 
When sampling methods such as Markov Chain Monte Carlo (MCMC; see e.g., \citet{Hogg2018-ApJS}) are employed, robust posteriors and uncertainties of parameters can be obtained \citep{Conroy2020-ApJS}. 
These methods require relatively high computational costs. 

The derivatives of a LC are expected to have considerable potential for probing the properties of a system. 
Typical signatures of eclipses can be observed in the derivatives \citep{IJspeert2021-AA, IJspeert2024-AA}. 
Classifying the LC derivatives of overcontact binaries allows us to infer their general properties \citep{Kouzuma2025-PASJ}. 
Binary parameters such as the mass ratio and orbital inclination can be estimated by measuring the time interval between two extrema found in the derivatives \citep{Kouzuma2023-ApJ, Kouzuma2025-PASJ-psaf111, Kouzuma2025-submit}. 
These parameter estimation methods can straightforwardly provide reasonable estimates and their associated uncertainties. 
In addition, this method requires no iterative processes and therefore involves low computational costs. 
This is helpful for quickly gaining insights into the properties of the system and for conducting subsequent detailed analyses. 
However, no straightforward method to estimate the fill-out factor has yet been reported. 

This paper proposes a simple method for estimating the fill-out factor of overcontact binaries using the derivatives of LCs. 
The proposed method enables us to estimate the fill-out factor from the key time interval measured between two local extrema in the derivatives of LCs. 
Section \ref{sec:data} introduces a sample of synthesized LCs for overcontact binary systems, which was used to identify associations between the fill-out factor and the key time intervals. 
On the basis of the identified association, we derive an empirical formula to estimate the fill-out factor in Section \ref{sec:method}. 
In Section \ref{sec:application}, we apply our method to real overcontact binary data and examine its effectiveness. 
Section \ref{sec:rd} shows the results and discusses the effectiveness.  
We summarize this work in Section \ref{sec:summary}.

\section{Data}\label{sec:data}
To derive an empirical formula for estimating the fill-out factor, we synthesized LCs of overcontact binaries using the PHOEBE 2.4 code \citep{Prsa2016-ApJS, Conroy2020-ApJS}. 
In this study, we adopted the following parameter ranges and steps: 
the mass ratio $q=M_\mathrm{s}/M_\mathrm{p}=0.05$--$0.95$ (in steps of 0.1), orbital inclination $i=30\tcdegree$--$90\tcdegree$ ($2\tcdegree$), primary and secondary star temperatures $T_\mathrm{p}$ ($T_\mathrm{s}$)$=4000$--$10000$ (1000) K, and fill-out factor $f=0.1$--$0.9$ (0.2). 
A fixed orbital period was assumed in this study. 
Preliminary tests using LC derivatives showed that varying the period had a negligible effect on the estimated binary parameters, justifying this simplification. 
The indices `p' and `s' refer to the primary and secondary, respectively; the primary star is defined as the more massive component of the binary. 
For systems with $f=0.1$ and $q=0.05$, PHOEBE was unable to generate a converged LC and terminated with a runtime error. 
A total of 74,431 LCs were generated for this work. 

Each synthesized LC was binned into 100 phase intervals. 
We also computed numerical derivatives of each LC with respect to time, up to the fourth order. 
The times of local extrema were measured by identifying zero slopes using linear interpolation between the derivatives at consecutive phase points.

\begin{figure}[ht]
 \begin{center}
  \includegraphics[width=0.75\textwidth]{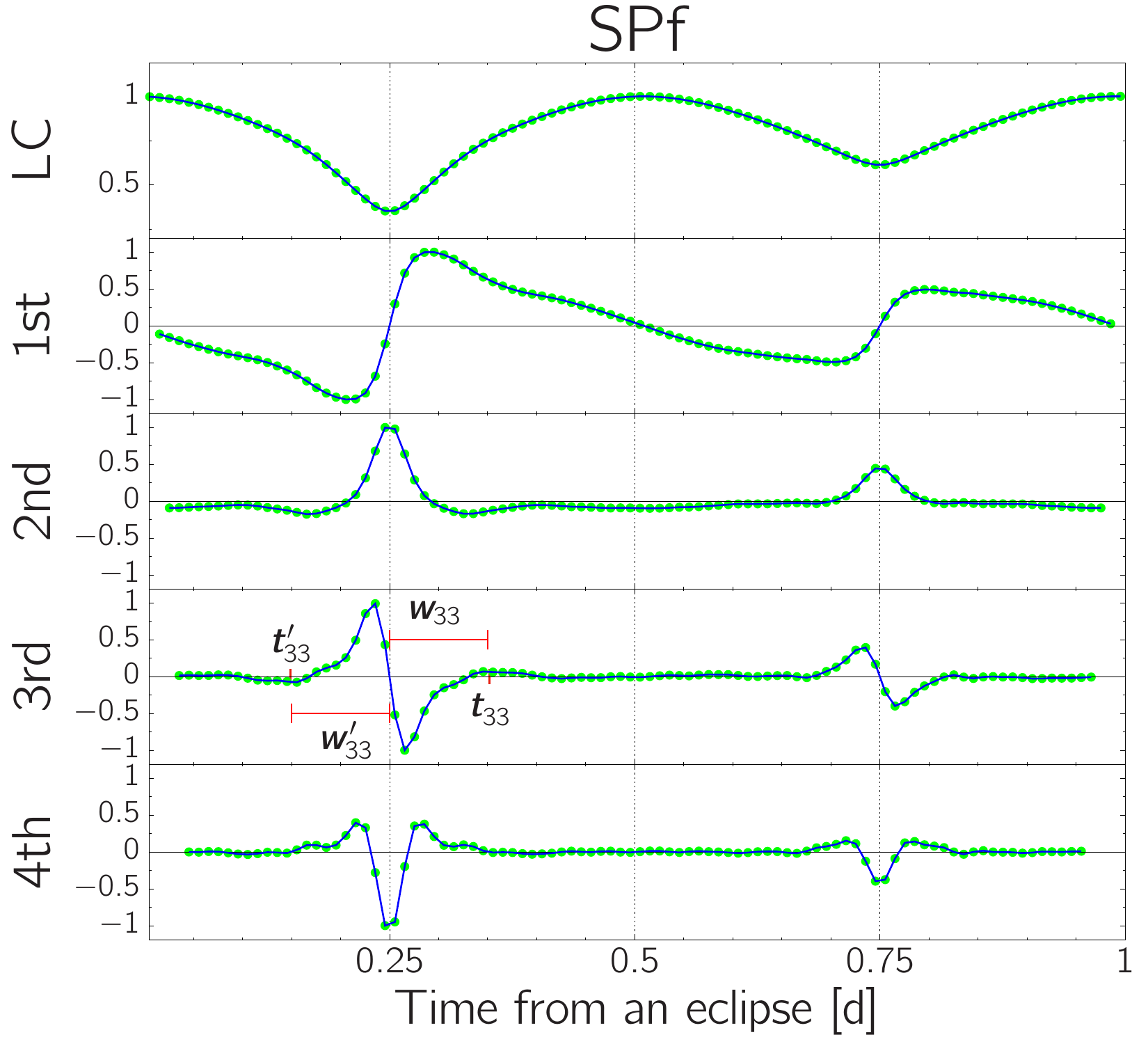}
 \end{center}
 \caption{Light curves and their first through fourth derivatives (from top to bottom) for a representative SPf-type LC. 
			The green points in the LC represent the average values of the synthesized LC in 100 bins, and the blue curve simply connects these points. 
			The derivatives were calculated using the LC averaged in 100 bins. 
			Each derivative is normalized to unity. 
			The key times used for estimating the fill-out factor and its associated uncertainty are labeled. 
 \label{fig:LCs}}
\end{figure}
\section{Method}\label{sec:method}
We first classified the synthesized LCs into five types (i.e., DP, SPp, SPb, SPf, and SPs), following the classification scheme introduced by \citet{Kouzuma2025-PASJ}, which describes the characteristic morphologies of LC derivatives. 
For each of the five types, we thoroughly examined the associations between the fill-out factors and all possible time intervals derived from two local extrema in the derivatives of LCs. 
As a result, for SPf-type LCs, we found several time intervals to be associated with the fill-out factors. 
Note that SPf systems are overcontact binaries with high mass ratios and high inclinations, and thus exhibit high eclipse obscurations \citep{Kouzuma2025-PASJ}. 
By comparing the accuracies of the fill-out factors estimated from each relationship, we finally determined that the following $W_f$ value is the most appropriate: 
\begin{equation}
	W_{f} = \frac{P}{t_{33} - t'_{33}} = \frac{P}{w_{33} + w'_{33}}, \label{eq:Wvalue-SPf}
\end{equation}
where $P$ is the orbital period, and $t_{ij}$ represents the time at either a local maximum or minimum around the primary (deeper) minima. 
Here, the index $i$ denotes the order of the derivative, and $j$ represents the index of the local extrema appearing near the eclipse time in the $i$-th derivative.
Note that $j=3$ is used to maintain consistency with the numbering adopted in previous studies \citep{Kouzuma2023-ApJ,Kouzuma2025-PASJ-psaf111,Kouzuma2025-submit}. 
We also define $w_{ij}$ as the time interval between $t_{ij}$ and the corresponding primary eclipse time. 
In Figure \ref{fig:LCs}, we show the times used in Equation (\ref{eq:Wvalue-SPf}). 

\begin{figure}[]
 \begin{center}
  \includegraphics[width=0.75\textwidth]{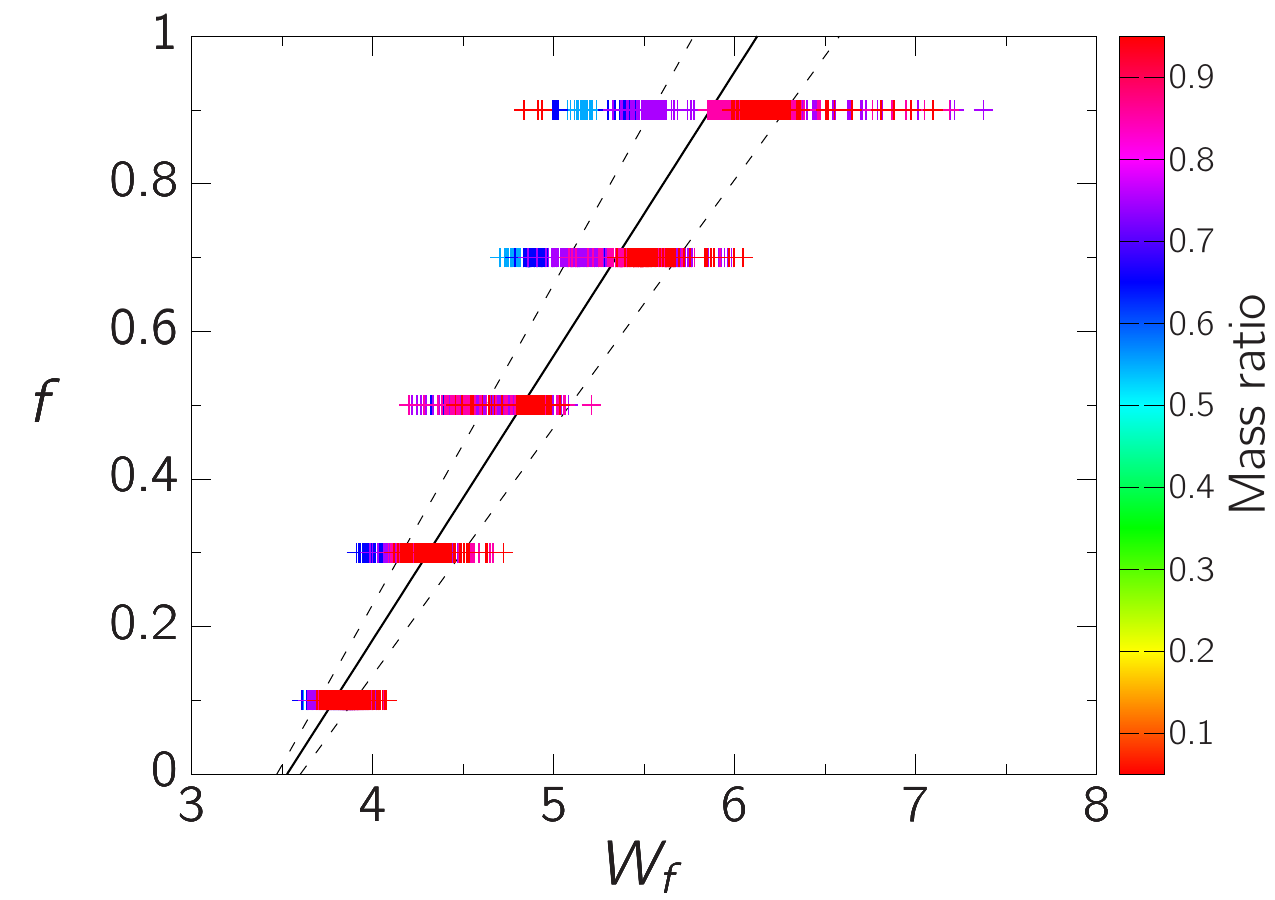}
 \end{center}
 \caption{Relationship between $W_f$ (Equation \ref{eq:Wvalue-SPf}) and the fill-out factor $f$. 
		  The solid line represents the regression line. 
		  The dashed lines show linear regression fits of the RMSEs, as expressed by Equation (\ref{eq:sigmaWf}), for each fill-out factor.
		  The color gradient represents the mass ratio, ranging from 0 (orange) to 1 (red). 
 \label{fig:W-f}}
\end{figure}

Figure \ref{fig:W-f} displays the relationship between the $W_f$ and $f$ for 3,850 SPf systems. 
The solid line represents a regression line that was fitted to minimize the sum of squares of the residuals in $W_f$. 
This regression line is expressed as follows: 
\begin{equation}
	f = 0.385 W_{f} - 1.359 \label{eq:f-estimator}. 
\end{equation}

In Figure \ref{fig:W-f}, a dispersion is observed in $W_f$ values for a given fill-out factor. 
The primary factor contributing to the dispersion is the difference in mass ratio, as seen in the figure. 
The standard deviations of the residuals in $W_f$ are $\sigma_{W_f}= 0.243$, corresponding to those for the fill-out factor of $\sigma_{f}= 0.093$. 
This dispersion contributes to the uncertainty of the estimated fill-out factor in this method. 

The uncertainties in the fill-out factor estimated with our method can be derived in the same manner as in \citet{Kouzuma2023-ApJ}, which considers two types of uncertainties: 
the measurement uncertainty in the timings of local extrema ($\delta W_f$), and the uncertainty in the empirical formula ($\sigma_{W_f}$). 

We assume that the uncertainty of the orbital period $P$ is sufficiently small to be negligible. 
Under the assumption that the two local extrema in the derivatives of a LC are perfectly symmetric (i.e., $w_{33}=w'_{33}$) and 
that the same value is measured twice, the variance $s^2$ is 
\begin{align}
	s^2 &= \frac{(w_{33}-\overline{w})^2 + (w'_{33}-\overline{w})^2}{2} \\
		&= \frac{\left(w_{33}-w'_{33}\right)^2}{4}, 
\end{align}
where 
\begin{equation}
	\overline{w} = \frac{w_{33}+w'_{33}}{2}. 
\end{equation}
Accordingly, the unbiased variance $s_u^2$ is calculated as 
\begin{equation}
	s_u^2 = 2 s^2. 
\end{equation}
Using Equation (\ref{eq:Wvalue-SPf}), the propagation of uncertainties yields 
\begin{equation}
\left(\delta W_f \right)^2 = \left(\frac{\partial W_f}{\partial w_{33}} \delta w_{33} \right)^2 + \left(\frac{\partial W_f}{\partial w'_{33}} \delta w'_{33} \right)^2. 
\end{equation}
Assuming that the two uncertainties (i.e., $\delta w_{33}$ and $\delta w'_{33}$) arise from measurement uncertainties and thus $\delta w_{33} = \delta w'_{33} = s_u$, we derive 
\begin{align}
	\delta W_f &= \frac{|w_{33}-w'_{33}|}{P} W_f^2. 
\end{align}

As seen in Figure \ref{fig:W-f}, the degree of dispersion increases with increasing fill-out factor. 
Taking this into account, we use the following equation as an uncertainty for $\sigma_{W_f}$, which was obtained by linearly fitting the root mean square errors (RMSEs) for each fill-out factor: 
\begin{equation}
	\sigma_{W_f} = 0.332 f + 0.064, \label{eq:sigmaWf}
\end{equation}
where $f$ is derived from Equation (\ref{eq:f-estimator}). 

We finally derive the uncertainty of the estimated fill-out factor as follows: 
\begin{align}
	\delta f &= 0.385 \sqrt{\sigma_{W_f}^2 + \delta W_f^2}. 
\end{align}
An expanded uncertainty is obtained through the multiplication of these values by an appropriate coverage factor. 

\section{Application to real binary data}\label{sec:application}
We applied the proposed method to real overcontact binaries to evaluate its effectiveness. 
We used the sample from the catalog compiled by \citet{Latkovic2021-ApJS}, which contains a total of 688 overcontact systems with entries of fundamental binary parameters, including the fill-out factor. 
These parameters were obtained in previous studies either through photometric analysis alone or through combined photometric and spectroscopic analyses. 
Hereafter, we refer to the former as the photometric sample and the latter as the spectroscopic sample. 

Using the Python package Lightkurve \citep{Lightkurve2018-code}, we extracted LCs from the TESS and Kepler archival data. 
The extracted LCs were phase-folded using the orbital periods determined via periodogram analysis. 
The phase was divided into 100 bins, and the data were averaged within each bin. 
When two or more LCs were available for a binary, the better-quality one was selected in such a manner that its derivatives were less noisy and smoother than the others. 
Apparent outliers in the LCs were removed in advance. 

After computing the LC derivatives up to the fourth order, we classified the LCs on the basis of \citet{Kouzuma2025-PASJ} and selected SPf-type LCs. 
Our final dataset comprised 4 (2 W-type and 2 A-type) spectroscopic and 19 (13 W-type and 6 A-type) photometric sample systems with extracted LCs. 
Here, W- and A-types are subtypes of W Ursae Majoris (W UMa) binaries introduced by \citet{Binnendijk1970-VA}. 
We then applied our proposed method to these LCs and estimated their fill-out factors together with their uncertainties.

\begin{figure}[]
 \begin{center}
  \includegraphics[width=0.75\textwidth]{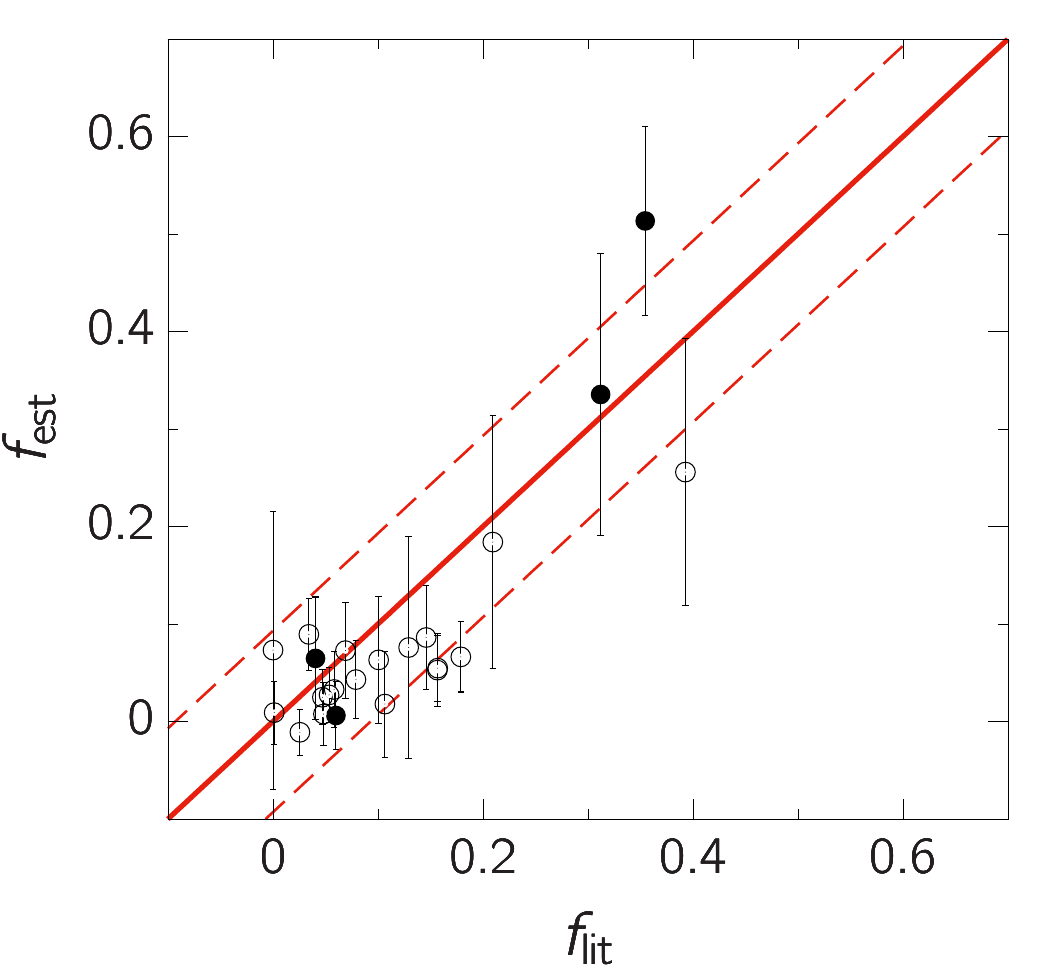}
 \end{center}
 \caption{Comparison between the fill-out factors estimated by the proposed method ($f_\text{est}$) and those reported in the literature ($f_\text{lit}$). 
		  The filled and open circles represent the fill-out factors of the spectroscopic and photometric sample LCs, respectively. 
		  The solid and dashed lines show $f_\text{est}=f_\text{lit}$ and $f_\text{est}=f_\text{lit} \pm 0.093$, respectively. 
 \label{fig:comparison}}
\end{figure}
\section{Results and Discussion}\label{sec:rd}
Figure \ref{fig:comparison} shows a comparison between our fill-out factor estimates ($f_\text{est}$) and corresponding literature values ($f_\text{lit}$) for the spectroscopic and photometric samples. 
The standard deviation of $(f_\text{est}-f_\text{lit})$ is 0.059, which is 0.034 smaller than that obtained for the synthesized LCs. 
This value is normally expected to be larger due mainly to observational errors that do not exist in the synthetic LCs, as noted in \citet{Kouzuma2023-ApJ, Kouzuma2025-PASJ-psaf111}. 
This unexpectedly small value is likely due to the fact that our sample systems have low fill-out factors, all of which are less than 0.5. 
As mentioned in Section \ref{sec:method}, in the synthesized sample LCs, the dispersion in $W_f$ increases with increasing fill-out factor. 
Indeed, when we compute the standard deviation for the subsample with $f<0.5$, we obtain 0.047, which is slightly smaller than 0.059, as expected from the aforementioned studies. 

The estimated fill-out factors for 57\% of the sample systems are consistent with their literature values within the estimated uncertainties. 
This percentage is approximately 11 percentage points smaller than the expected value (i.e., $\sim 68$\%) based on the standard uncertainty. 
In this analysis, we did not take into account the uncertainties of the literature values, because they are often not explicitly stated in the original papers. 
Even when reported, these uncertainties were likely to be unrealistically small in most cases. 
If more realistic uncertainties for the literature values were available, the observed percentage (i.e., 57\%) would likely approach the expected one. 
Specifically, assuming the uncertainties of $f_\text{lit}$ to be approximately 11–17\% of their values, about 68\% of the estimated fill-out factors agree with the literature values. 
Consequently, our method can be considered sufficiently practical.

\section{Summary}\label{sec:summary}
We have proposed a new method for estimating the fill-out factors of overcontact binaries. 
Our method is applicable to the SPf-type LCs, which are specified using the classification scheme for the derivatives of LCs proposed by \citet{Kouzuma2025-PASJ}. 
Once the LC derivatives are obtained, the fill-out factor can be estimated simply by measuring the time interval between two local extrema in the third derivative and applying the empirical formula. 
No iterative process is required. 
The application to real overcontact binary data demonstrated that the proposed method can reasonably estimate reliable fill-out factors and their associated uncertainties. 

This method is also expected to yield reliable estimates even when a third body or starspots are present in the binary system, provided that they do not exhibit significant brightness variations, as demonstrated in \citet{Kouzuma2023-ApJ,Kouzuma2025-PASJ-psaf111}. 
However, as mentioned in \citet{Kouzuma2025-submit}, if starspots were to affect the LC enough to change the relative depths of the minima and thus swap the primary and secondary minima, the fill-out factor estimated by our method may differ significantly from the true value. 

While our method relies on LC derivatives and does not employ machine learning, machine learning algorithms could in principle be applied to the same problem. 
In particular, applying machine learning to the derivatives of LCs may provide more reliable estimates of binary parameters than applying it directly to the LCs themselves.

In combination with the methods for estimating the mass ratio and inclination proposed by \citet{Kouzuma2023-ApJ,Kouzuma2025-PASJ-psaf111,Kouzuma2025-submit}, applying the present method to estimate the fill-out factor would enable a deeper understanding of the system’s properties. 
Furthermore, obtaining fill-out factors for a large number of systems would also contribute to statistical studies.

\vspace{6pt} 

\funding{This work was supported by JSPS KAKENHI Grant Number 25K07358. }

\acknowledgments{
The author would like to thank the anonymous referees for helpful comments and suggestions which improved the paper. 
This paper includes data collected by the TESS mission. Funding for the TESS mission is provided by the NASA’s Science Mission Directorate. 
This paper includes data collected by the Kepler mission and obtained from the MAST data archive at the Space Telescope Science Institute (STScI). 
Funding for the Kepler mission is provided by the NASA Science Mission Directorate. 
STScI is operated by the Association of Universities for Research in Astronomy, Inc., under NASA contract NAS 5–26555. 
This research made use of Lightkurve, a Python package for Kepler and TESS data analysis (Lightkurve Collaboration, 2018). 
}

\conflictsofinterest{The authors declare no conflicts of interest. }

\begin{adjustwidth}{-\extralength}{0cm}

\reftitle{References}

\def\DeclareAbbreviation#1#2{%
   \DeclareRobustCommand*#1{\@journalname{#2}}}
\def\@journalname#1{{\normalfont#1}}
\DeclareAbbreviation\aj{AJ}
\DeclareAbbreviation\araa{ARA\&A}
\DeclareAbbreviation\apj{ApJ}
\DeclareAbbreviation\apjl{ApJL}
\DeclareAbbreviation\apjs{ApJS}
\DeclareAbbreviation\ao{Appl.\ Opt.}
\DeclareAbbreviation\apss{Ap\&SS}
\DeclareAbbreviation\aap{A\&A}
\DeclareAbbreviation\aapr{A\&AR}
\DeclareAbbreviation\aaps{A\&AS}
\DeclareAbbreviation\azh{AZh}
\DeclareAbbreviation\baas{BAAS}
\DeclareAbbreviation\icarus{ICARUS}
\DeclareAbbreviation\jrasc{JRASC}
\DeclareAbbreviation\memras{MmRAS}
\DeclareAbbreviation\mnras{MNRAS}
\DeclareAbbreviation\pra{Phys.\ Rev.\ A}
\DeclareAbbreviation\prb{Phys.\ Rev.\ B}
\DeclareAbbreviation\prc{Phys.\ Rev.\ C}
\DeclareAbbreviation\prd{Phys.\ Rev.\ D}
\DeclareAbbreviation\pre{Phys.\ Rev.\ E}
\DeclareAbbreviation\prl{Phys.\ Rev.\ Lett.}
\DeclareAbbreviation\pasp{PASP}
\DeclareAbbreviation\pasj{PASJ}
\DeclareAbbreviation\qjras{QJRAS}
\DeclareAbbreviation\raa{Res.\ Astron.\ Astronphys.}
\DeclareAbbreviation\skytel{S\&T}
\DeclareAbbreviation\solphys{Sol.\ Phys.}
\DeclareAbbreviation\sovast{Soviet\ Ast.}
\DeclareAbbreviation\ssr{Space\ Sci.\ Rev.}
\DeclareAbbreviation\zap{ZAp}
\DeclareAbbreviation\nat{Nature}
\DeclareAbbreviation\iaucirc{IAU\ Circ.}
\DeclareAbbreviation\aplett{Astrophys.\ Lett.}
\DeclareAbbreviation\apspr{Astrophys.\ Space\ Phys.\ Res.}
\DeclareAbbreviation\bain{Bull.\ Astron.\ Inst.\ Netherlands}
\DeclareAbbreviation\fcp{Fund.\ Cosmic\ Phys.}
\DeclareAbbreviation\gca{Geochim.\ Cosmochim.\ Acta}
\DeclareAbbreviation\grl{Geophys.\ Res.\ Lett.}
\DeclareAbbreviation\jcp{J.\ Chem.\ Phys.}
\DeclareAbbreviation\jgr{J.\ Geophys.\ Res.}
\DeclareAbbreviation\jqsrt{J.\ Quant.\ Spectrosc.\ Radiat.\ Transfer}
\DeclareAbbreviation\memsai{Mem.\ Soc.\ Astron.\ Italiana}
\DeclareAbbreviation\nphysa{Nucl.\ Phys.\ A}
\DeclareAbbreviation\physrep{Phys.\ Rep.}
\DeclareAbbreviation\physscr{Phys.\ Scr.}
\DeclareAbbreviation\planss{Planet.\ Space\ Sci.}
\DeclareAbbreviation\procspie{Proc.\ SPIE}
\DeclareAbbreviation\aip{AIP Conf.\ Proc.}
\DeclareAbbreviation\asp{ASP Conf.\ Ser.}
\DeclareAbbreviation\actaa{AcA}

\bibliographystyle{Definitions/mdpi}

\PublishersNote{}
\end{adjustwidth}
\end{document}